\documentclass[resetfootnote,twocolumn]{aastex701}

\newcommand{\xmm}{{\em XMM-Newton}}
\newcommand{\swift}{{\em Swift}}
\newcommand{\suz}{{\em Suzaku}}
\newcommand{\src}{BG\,CMi}

\received{XXXX}
\revised{XXXX}
\accepted{XXXX}
\submitjournal{ApJ}
\graphicspath{{./}{Figures/}}

\begin{document}

\title{A multi-wavelength study of the 2025 low state of the intermediate polar BG\,CMi}

\author[orcid=0000-0002-8808-520X]{A. W. Shaw}
\affiliation{Department of Physics \& Astronomy, Butler University, 4600 Sunset Avenue, Indianapolis, IN 46208, USA}
\email[show]{awshaw@butler.edu}  

\author[orcid=0000-0002-8286-8094]{K. Mukai}
\affiliation{CRESST II and X-ray Astrophysics Laboratory, NASA/GSFC, Greenbelt, MD 20771, USA}
\affiliation{Department of Physics, University of Maryland, Baltimore County, 1000 Hilltop Circle, Baltimore, MD 21250, USA}
\email{koji.mukai-1@nasa.gov} 

\author[orcid=0000-0003-3944-6109]{C. O. Heinke}
\affiliation{Department of Physics, University of Alberta, Edmonton, AB, T6G 2E1, Canada}
\email{heinke@ualberta.ca}   

\author{C. G. Nixon}
\affiliation{Department of Physics \& Astronomy, Butler University, 4600 Sunset Avenue, Indianapolis, IN 46208, USA}
\email{cgnixon@butler.edu}


\author[orcid=0000-0002-7004-9956]{D. A. H. Buckley}
\affiliation{South African Astronomical Observatory, PO Box 9, Observatory, 7935, Cape Town, South Africa}
\affiliation{Department of Astronomy, University of Cape Town, Private Bag X3, Rondebosch 7701, South Africa}
\affiliation{Department of Physics, University of the Free State, PO Box 339, Bloemfontein 9300, South Africa}
\email{dah.buckley@saao.nrf.ac.za}

\author[orcid=0000-0001-5541-2836]{P. A Dubovsk\'{y}}
\affiliation{Vihorlat Observatory in Humenn\'e, Slovakia}
\affiliation{Variable Star Section, Slovak Astronomical Society}
\email{var@kozmos.sk}
\author[orcid=0000-0003-0125-8700]{F.-J. Hambsch}
\affiliation{American Association of Variable Star Observers, 185 Alewife Brook Parkway, Suite 410, Cambridge, MA 02138, USA}
\affiliation{Groupe Européen d’Observations Stellaires (GEOS), 23 Parc de Levesville, 28300 Bailleau l’Evêque, France}
\affiliation{Bundesdeutsche Arbeitsgemeinschaft für Veränderliche Sterne (BAV), Munsterdamm 90, 12169 Berlin, Germany}
\affiliation{Vereniging Voor Sterrenkunde (VVS), Zeeweg 96, 8200 Brugge, Belgium}
\email{hambsch@telenet.be}

\author{J. Hilburn}
\affiliation{Great Basin Observatory, Great Basin National Park, Baker, NV 89311}
\email{jerry@epsilonorion.com}

\author{K.\,Petr\'{i}k}
\affiliation{Observatory and Planetarium M. R. \v{S}tef\'{a}nik in Hlohovec, Slovakia}
\email{kpetrik@hvezdaren.org}

\author[orcid=0000-0002-7092-0326]{R. M. Plotkin}
\affiliation{Department of Physics, University of Nevada, Reno, NV 89557, USA}
\affiliation{Nevada Center for Astrophysics, University of Nevada, Las Vegas, NV 89154, USA}
\email{rplotkin@unr.edu}

\author[orcid=0000-0002-5956-2249]{S. B. Potter}
\affiliation{South African Astronomical Observatory, PO Box 9, Observatory, 7935, Cape Town, South Africa}
\affiliation{Department of Physics, University of Johannesburg, PO Box 524, Auckland Park 2006, South Africa}
\email{sbp@saao.ac.za}

\author{N. Rawat}
\affiliation{South African Astronomical Observatory, PO Box 9, Observatory, 7935, Cape Town, South Africa}
\email{rawatn@saao.ac.za}

\author[orcid=0000-0003-1331-5442]{T. Shahbaz}
\affiliation{Instituto de Astrof\'{i}sica de Canarias, E-38205 La Laguna, Tenerife, Spain}
\affiliation{Departamento de Astrof\'{i}sica, Universidad de La Laguna, E-38206 La Laguna, Tenerife, Spain}
\email{tariqshahbaz1@gmail.com}


\author{S. Dufoer}
\affiliation{American Association of Variable Star Observers, 185 Alewife Brook Parkway, Suite 410, Cambridge, MA 02138, USA}
\affiliation{Vereniging Voor Sterrenkunde (VVS), Zeeweg 96, 8200 Brugge, Belgium}
\email{sdufoer@gmail.com}

\author{S. Dvorak}
\affiliation{American Association of Variable Star Observers, 185 Alewife Brook Parkway, Suite 410, Cambridge, MA 02138, USA}
\affiliation{Rollinghills Observatory, Clermont, FL 34711, USA}
\email{sdvorak@rollinghillsobs.org}

\author{D. Messier}
\affiliation{American Association of Variable Star Observers, 185 Alewife Brook Parkway, Suite 410, Cambridge, MA 02138, USA}
\affiliation{Center for Backyard Astrophysics, 35 Sergeants Way, Lisbon, CT  06351}
\email{dpmessier@aol.com}

\author{G. Myers}
\affiliation{American Association of Variable Star Observers, 185 Alewife Brook Parkway, Suite 410, Cambridge, MA 02138, USA}
\email{gordonmyers@hotmail.com}

\author{P. Nelson}
\affiliation{American Association of Variable Star Observers, 185 Alewife Brook Parkway, Suite 410, Cambridge, MA 02138, USA}
\affiliation{Wild Cherry Observatory, Wild Cherry  Road,Lockwood South, Victoria 3551, Australia}
\email{pnelson@aussiebb.com.au}

\author{R. Sabo}
\affiliation{American Association of Variable Star Observers, 185 Alewife Brook Parkway, Suite 410, Cambridge, MA 02138, USA}
\affiliation{1344 Post Drive, Bozeman, MT 59715, USA}
\email{rsabo333@gmail.com}

\author{J. Ulowetz}
\affiliation{American Association of Variable Star Observers, 185 Alewife Brook Parkway, Suite 410, Cambridge, MA 02138, USA}
\affiliation{Center for Backyard Astrophysics, 855 Fair Ln, Northbrook, IL 60062, USA}
\email{joe700a@gmail.com}

\author{T. Vanmunster}
\affiliation{American Association of Variable Star Observers, 185 Alewife Brook Parkway, Suite 410, Cambridge, MA 02138, USA}
\affiliation{CBA Belgium Observatory, Walhostraat 1A, B-3401 Landen, Belgium}
\affiliation{CBA Extremadura Observatory, E-06340 Fregenal de la Sierra, Badajoz, Spain}
\affiliation{Vereniging Voor Sterrenkunde (VVS), Zeeweg 96, 8200 Brugge, Belgium}
\email{tonny.vanmunster@gmail.com}


\begin{abstract}

We present multi-wavelength observations of the first recorded low state of the intermediate polar \src. Optical monitoring of the source by members of the American Association of Variable Star Observers reveals a decrease of $\sim0.5$ mag that lasted $\sim 50$ d in early 2025. During the low state the optical timing properties imply that \src\ underwent a change in the accretion mode, as power at the spin frequency $\omega$ dramatically dropped. An \xmm\ observation revealed a substantial decrease in intrinsic absorption and a slight increase in intrinsic X-ray luminosity, compared to archival \suz\ data. Timing analysis of the X-ray light curves shows that power shifted from the orbital frequency $\Omega$ (prominent in \suz\ data) to $2\Omega$ in the low state \xmm\ data, along with the strengthening of certain orbital sidebands. We suggest that \src\ transitioned to disk-overflow accretion, where the white dwarf accreted matter via both a disk and a stream, the latter becoming more dominant during the low state due to a decrease in the mass and size of the disk.

\end{abstract}





\section{Introduction} 
\label{sec:intro}

Intermediate Polars (IPs) are a class of accreting white dwarfs and a subclass of cataclysmic variables in which a white dwarf (WD) accretes matter from a (typically) main-sequence companion. In IPs, a partial accretion disk 
can 
form. However the magnetic field of the WD is strong enough ($B\gtrsim10^6$ G) to disrupt the innermost portion of the disk, such that the innermost edge lies at the magnetospheric radius. Material at the innermost edge of the disk flows along the magnetic field lines onto the magnetic poles of the WD \citep[see e.g.][for reviews]{Patterson-1994,Ferrario-2015,Ferrario-2020}. 

One of the key observational signatures of IPs is multi-wavelength periodic variability. Unlike in polars, another class of magnetic CV, the spin period of the WD ($P_{\rm spin}$) and the binary orbital period ($P_{\rm orb})$ are not synchronized in IPs. Consequently, the power spectra of IP light curves at optical, UV, and X-ray wavelengths are typically dominated by the WD spin frequency ($\omega=1/P_{\rm spin}$), the orbital frequency ($\Omega=1/P_{\rm orb}$) and the beat frequency ($\omega - \Omega$). IPs have also been known to exhibit variability at orbital sidebands such as $\omega+\Omega$ and $\omega-2\Omega$ \citep{Warner-1986}.

The timing properties of IPs contain information about the accretion properties of the system. Though the standard picture of accretion in IPs is one of a partial disk that is truncated at the magnetospheric radius (the `disk-fed' model), there is strong evidence that at least one IP, V2400\,Oph, has accreted via a `stream-fed' mode in which material travels along a stream toward the WD magnetosphere without forming a disk \citep[see e.g.][]{Buckley-1995,Buckley-1997,Hellier-2002a}. A third accretion mode known as `disk-overflow' can be considered as a hybrid of stream- and disk-fed accretion modes, as both types are observed simultaneously \citep{Hellier-1993,Armitage-1996,Armitage-1998}. IPs have shown evidence of switching between accretion modes, in fact V2400\,Oph itself, despite being the canonical `diskless IP', has exhibited signs of disk-overflow accretion \citep{Joshi-2019}, while a recent study has revealed a much more complex accretion system than previously thought \citep{Langford-2022}. The observational evidence for these changes in accretion modes lies in the power spectra of IP light curves.

\citet{Wynn-1992} and \citet{Ferrario-1999} have shown that the strengths of the different peaks in IP power spectra can be used to determine the dominant accretion mode in the system. For example, a typical disk-fed IP will show strong variability on $\omega$ at optical/UV and X-ray wavelengths, owing to the presence of a Keplerian disk, the inner edge of which co-rotates with the WD. However, in the stream-fed regime, the power spectra start to become more complex. \citet{Ferrario-1999} showed that, for diskless IPs, the X-ray light curves might show variability at $\omega$, $\omega-\Omega$ and $2\omega-\Omega$, with the strength of the peaks in the power spectra varying with binary inclination, $i$. In the optical, the power spectrum of a diskless IP might show strong peaks at $\omega$, $\omega-2\Omega$ and $2(\omega-\Omega)$, depending on $i$, with the power spectrum of a model low-inclination system ($i\sim20^\circ$) completely dominated by the $2(\omega-\Omega)$ beat. 

Changes in IP accretion modes have been seen to be coincident with a decrease in a source's optical flux. These so-called `low-states' have been known to exist for a while \citep[e.g.][]{Garnavich-1988}, but have been relatively poorly studied until recently. The IP FO\,Aqr has exhibited three excursions to a low state since 2015, during which it transitioned from a disk-fed to a stream-fed or disk-overflow accretion mode \citep[e.g.]{Kennedy-2017,Littlefield-2020}. \citet{Covington-2022} found evidence for changing accretion geometry in two IPs during their low states. In DO\,Dra, accretion may even have completely stopped for a short period during its low state \citep{Hill-2022}. It is becoming evident that IP low states and changes in accretion geometry are linked. However, to truly understand the accretion processes at work, one must take a multi-wavelength approach. Only one IP low state has been studied at X-ray energies \citep[FO\,Aqr;][]{Kennedy-2017}, revealing changes in the X-ray spectrum that were interpreted as a switch to a stream-fed geometry. X-ray observations allow us to study the shock close to the surface of the WD, while optical photons typically originate from the accretion disk, including the hotspot, as well as from the irradiated face of the secondary. Multi-wavelength observations allow us to paint a more complete picture of the physical processes at work during a low state.

Here, we present only the second deep X-ray study of an IP low state, after FO\,Aqr \citep{Kennedy-2017}. In this work, we use the practical definition of ``low state" as defined by \citet{Covington-2022}, i.e. a sustained drop in optical flux of $\gtrsim0.5$ mag from the average. However, we acknowledge that there could be many meaningful definitions of a low state \citep[see e.g.][]{Garnavich-1988, Kennedy-2017, Littlefield-2020}.

We performed observations of the IP \src\ with \xmm\ in March 2025 during a time when the optical flux had decreased significantly. The X-ray observations were performed simultaneously with optical observations in a number of wavebands. The paper is structured as follows: in Section \ref{sec:obs} we detail the observations and data reduction, in Section \ref{sec:Analysis} we present both spectral and temporal analysis of the data and finally we discuss our results and present conclusions in Sections \ref{sec:discussion} and \ref{sec:conclusions}, respectively.

\subsection{BG\,CMi}

BG Canis Minoris (\src) was first reported as the X-ray source 3A\,0729+103 and quickly associated with a newly discovered CV  \citep{McHardy-1982}, before being confirmed as an IP \citep{McHardy-1984}. The WD has a spin period of $P_{\rm spin}=913.5$ s \citep{Hellier-1997} which has been shown to spin up over many observing campaigns, though the rate of spin-up has slowed down \citep{Kim-2005,Bonnardeau-2016,Patterson-2020}. The binary orbital period is $P_{\rm orb}=3.23$ h \citep{Bonnardeau-2016} such that $P_{\rm spin}/P_{\rm orb}\sim0.08$ and thus \src\ is expected to have a partial accretion disk, assuming spin equilibrium \citep{King-1991}. In March 2025, \src\ was reported to have entered a low state, with optical monitoring of the source noting an overall fading of $\sim0.5$ magnitudes \citep{Littlefield-2025}. Prior to this, no low states of the source had been reported. 

\section{Observations and Data Reduction}
\label{sec:obs}

\subsection{XMM-Newton}

\src\ was observed by \xmm\ for a total of 31\,ks on 2025 March 23 at 19:00 UTC. The European Photon Imaging Camera (EPIC) detectors, consisting of the EPIC-pn \citep{Struder-2001} and the two EPIC-MOS \citep{Turner-2001} were all operated in Large Window mode, affording 47.7 ms time resolution in pn and 0.9 s time resolution in MOS. The EPIC instruments also utilized a thin optical blocking filter. The optical/UV monitor \citep[OM;][]{Mason-2001} provided simultaneous observations in the $V$-band and was operated in Fast mode, providing 0.5 s time resolution. 

Data were reduced using the \xmm\ Science Analysis System ({\sc sas}) v22.1.0. To reprocess the EPIC data we used {\tt epproc} and {\tt emproc} for pn and MOS, respectively. EPIC event lists were filtered for periods of high background flaring activity. Photon arrival times in the cleaned event files were corrected to the solar system barycenter using the SAS task {\tt barycen}. We then extracted light curves in the 0.2--10 keV band for all EPIC detectors. For EPIC-pn we extracted a source light curve from a circular region of radius $40\arcsec$ centered on the source. Background photons were extracted from a circular region of radius $60\arcsec$ centered on a source-free region on the same chip as the source.

\begin{figure*}
    \centering
    \includegraphics[width=0.95\textwidth]{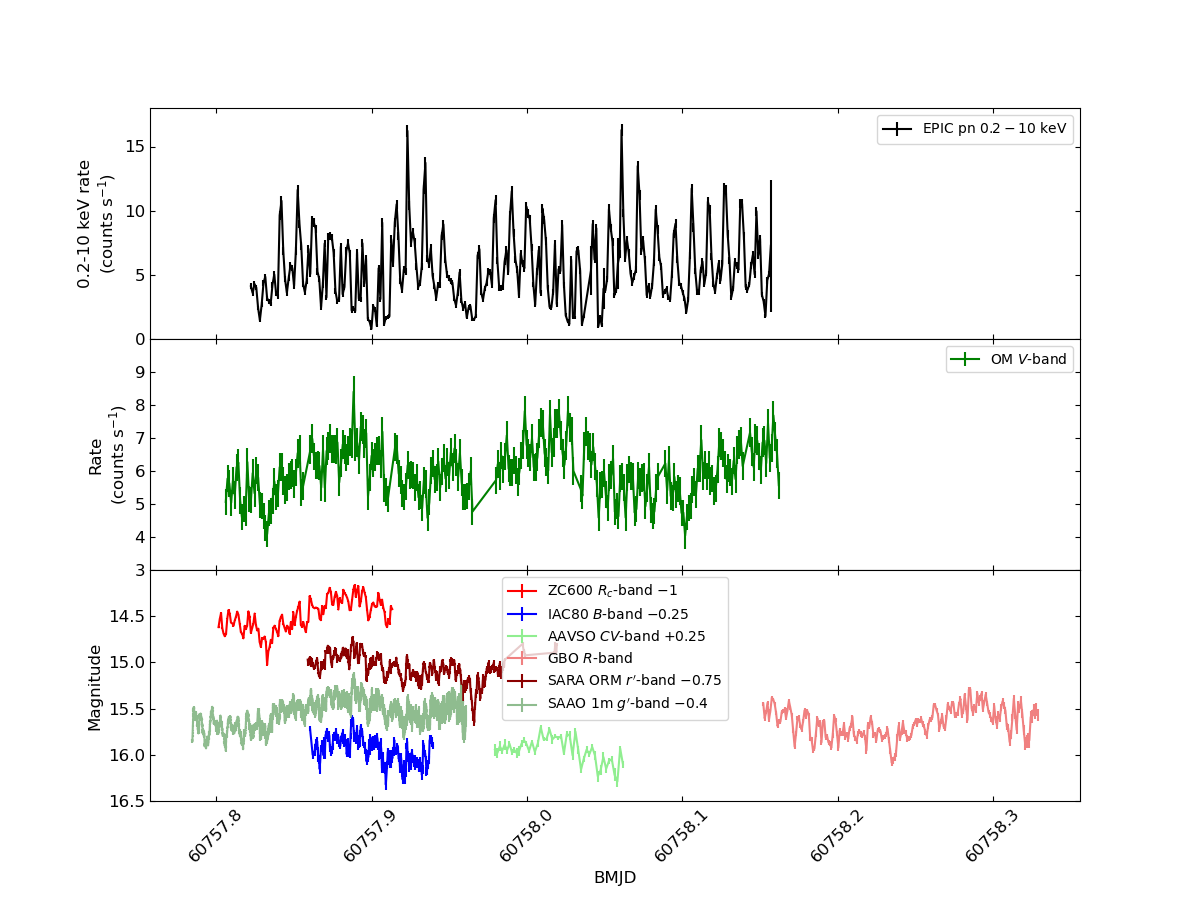}
    \caption{{\em Upper Panel}: \xmm/EPIC-pn 0.2--10 keV light curve of \src with 100 s time resolution. {\em Middle Panel}: \xmm/OM $V$-band light curve of \src with 100 s time resolution. {\em :Lower Panel}: Ground-based optical light curves of \src. Filters and telescopes are denoted in the legend. Some light curves have been offset for clarity. All light curves have been corrected to the solar system barycenter and thus times are given in Barycentric Modified Julian Date (BMJD).}
    \label{fig:XMMcampaign_LC}
\end{figure*}

The 0.2--10 keV EPIC-pn light curve is shown in the upper panel of Fig.\ \ref{fig:XMMcampaign_LC} and shows large amplitude variability. The count rate reaches upwards of 15 counts s$^{-1}$, meaning we must consider the effects of photon pile-up in the spectra.\footnote{The maximum rate for the EPIC-pn to avoid deteriorated response due to photon pile-up in Large Window mode is 3 counts s$^{-1}$: \href{https://xmm-tools.cosmos.esa.int/external/xmm_user_support/documentation/uhb/epicmode.html}{https://xmm-tools.cosmos.esa.int/external/xmm\_user\_support/\\documentation/uhb/epicmode.html}} The MOS light curves are not shown here, but the count rate reaches upwards of 4 counts s$^{-1}$, so we must also consider pile-up when extracting the MOS spectra. 

For EPIC-pn, the amplitude of the variability is 
quite large. 
To develop a pile-up correction with minimal photon losses, we split our cleaned, barycentered event file into three new event files filtered on source count rate. The first contains 
times 
when the source count rate is $<3$ counts s$^{-1}$, the second contains 
times 
when the source rate is in the range $3-10$ counts s$^{-1}$, and the third when the source rate is $>10$ counts s$^{-1}$ (i.e. the strongest pile-up). We 
examined the pile-up levels using the {\sc sas} task {\tt epatplot}. We then extracted source spectra from each event file, using a circular region of radius $40\arcsec$ for the $<3$ counts s$^{-1}$ event file, an annulus of inner radius $10\arcsec$ and outer radius $40\arcsec$ for the $3-10$ counts s$^{-1}$ event file, and an annulus of inner radius $12\farcs5$ and outer radius $40\arcsec$ for the $>10$ counts s$^{-1}$ event file. For the highest rate event file, we also only considered single events ({\sc PATTERN==0}), which are less susceptible to pile-up. Background spectra were extracted from a circular region of radius $60\arcsec$ centered on a source-free region on the same chip. 
Response matrices and ancillary responses were generated using the {\sc sas} tools {\tt rmfgen} and {\tt arfgen}, respectively. 
The three count-rate selected spectra were 
combined using {\tt epicspeccombine} to produce a pile-up corrected spectrum. The pn spectrum has been grouped such that each bin has a minimum of 20 counts. A caveat of using {\tt epicspeccombine} is that the tool assumes that the source regions are exactly the same for all co-added spectra. However, this is clearly not the case here, 
as we utilized different extraction regions depending on the count rate. To remedy this, we include a cross-normalization constant for all spectral fits simultaneous with MOS. As we will see in Section \ref{sec:XMMspec}, the spectral shape is consistent across all three EPIC instruments, but the flux is overestimated in the EPIC-pn spectrum and corrected by the cross-normalization constant.

For MOS1 we extracted a source spectrum from an annulus of inner radius $7\farcs5$ and outer radius $30\arcsec$ centered on the source and a background spectrum from a circular region of radius $50\arcsec$ centered on a source-free region on the same chip as the source. For MOS2 we extracted a source spectrum from an annulus of inner radius $12\farcs5$ and outer radius $30\arcsec$ centered on the source and a background spectrum from a circular region of radius $40\arcsec$ centered on a source-free region on the same chip as the source. Response matrices and ancillary responses were generated using the {\sc sas} tools {\tt rmfgen} and {\tt arfgen}, respectively. Spectra from all EPIC instruments have been grouped such that each bin has a minimum of 20 counts.

\subsection{Ground-based optical observations}

We organized a dedicated ground-based optical observing campaign simultaneous with the \xmm\ observations, which we describe here chronologically. Starting on 2025 March 23 at 18:25 UTC, we observed \src\ unfiltered and in the $g'$-band with the 1 m Telescope of the South African Astronomical Observatory (SAAO). We observed the source for approximately 4.5 h with exposure times of 5 s. All exposures were GPS triggered. Master bias frames were created and applied. Data reductions were performed using the {\sc python} packages {\tt photutils} v2.2.0 \citep{Bradley-2025}, Astropy \citep{Astropy-2013,Astropy-2018,Astropy-2022} and {\tt astroquery.gaia}. Calibrated magnitudes from aperture photometry were calculated using 2 field sources present in the European Space Agency {\em Gaia} Archive\footnote{\url{https://gea.esac.esa.int/archive/}}.

On 2025 March 23 at 19:13 UTC, we observed \src\ using the the Csere Telescope 600/2500 mm at M.R. \v{S}tef\'{a}nik Observatory in Hlohovec, Slovakia using the Atik 383L CCD Camera and the $R_c$ filter. The exposure time was 60 s, and the surce was observed for approximately 2 h 43 min. The data were reduced in a standard way using the DAOPHOT package in {\sc iraf}\footnote{\url{https://github.com/iraf-community/iraf}}, custom-written scripts, and FORTRAN programs created by T. Pribulla (priv. comm.). Initially, master darks and flats were produced. Then all object frames were cleaned from bad pixels and photometrically calibrated. All frames were astrometrically solved to define the pixel to world coordinate system (WCS) transformation, and finally, aperture photometry was performed.

On 2025 March 23 at 20:36 UTC, time-resolved optical photometry of \src\ was carried out using the CAMELOT-2  instrument on the  82 cm IAC80 telescope in Teide, Tenerife. We used the Johnson $B$ ﬁlter with an exposure time of 120\,s and obtained a total of 188 images. First, the CAMELOT-2 pipeline software was used to de-bias and flat-field the data. Then the HiPERCAM pipeline software\footnote{\url{https://github.com/HiPERCAM/hipercam}} was used to extract the target count rates using aperture photometry with a seeing-dependent circular aperture tracking the centroid of the source. The sky background was computed using the clipped mean of an annular region around the target and relative photometry of \src\ was carried out with respect to a local standard star
(\citealt{Henden-1995}; BG CMi-5, with B=13.16).

Almost simultaneously, we observed \src\ in the $r'$-band with the 1 m Jacobus Kapteyn Telescope (also known as SARA-ORM) at the Roque de los Muchachos, La Palma, Spain, one of the remote facilities operated by the Southeastern Association for Research in Astronomy \citep[SARA;][]{Keel-2017}. We observed the source for approximately 3 h 51 min with exposure times of 20 s. Master calibrations consisting of bias frames and flat fields were created using the Astropy-affiliated {\sc python} package {\tt ccdproc} \citep{Craig-2017} and applied to the raw science images. The calibrated science images were then aligned in WCS using the {\tt astrometry\_net} module included with the Astropy-affiliated {\tt astroquery} {\sc python} package \citep{Ginsburg-2019}. Aperture photometry was performed using {\tt photutils} v2.2.0 \citep{Bradley-2025}. Calibrated magnitudes were calculated using 7 field sources present in the Pan-STARRS1 catalog \citep{Chambers-2016,Magnier-2020}.

Starting at 2025 March 24 at 03:38 UTC, \src\ was also observed with the 0.7 m robotic Great Basin Observatory (GBO) in Great Basin National Park, Nevada, USA for approximately 4 h 15 min, overlapping with the very end of the \xmm\ observation. Observations were performed in the $R$-band, with an exposure time of 60 s per image. Data reduction and plate solving was performed in the same way as for the SARA-ORM images, and aperture photometry was performed using {\tt photutils}. As there were no field stars with calibrated $R$-band magnitudes, we chose 8 field stars with known $r'$- and $i'$-band magnitudes from the Pan-STARRS1 catalog and converted them to $R$-band magnitudes using the \citet{Jordi-2006} color transformations between photometric systems. We then used the resultant $R$-band magnitudes to calibrate the photometry of \src.

\begin{figure*}
    \centering
    \includegraphics[width=0.95\textwidth]{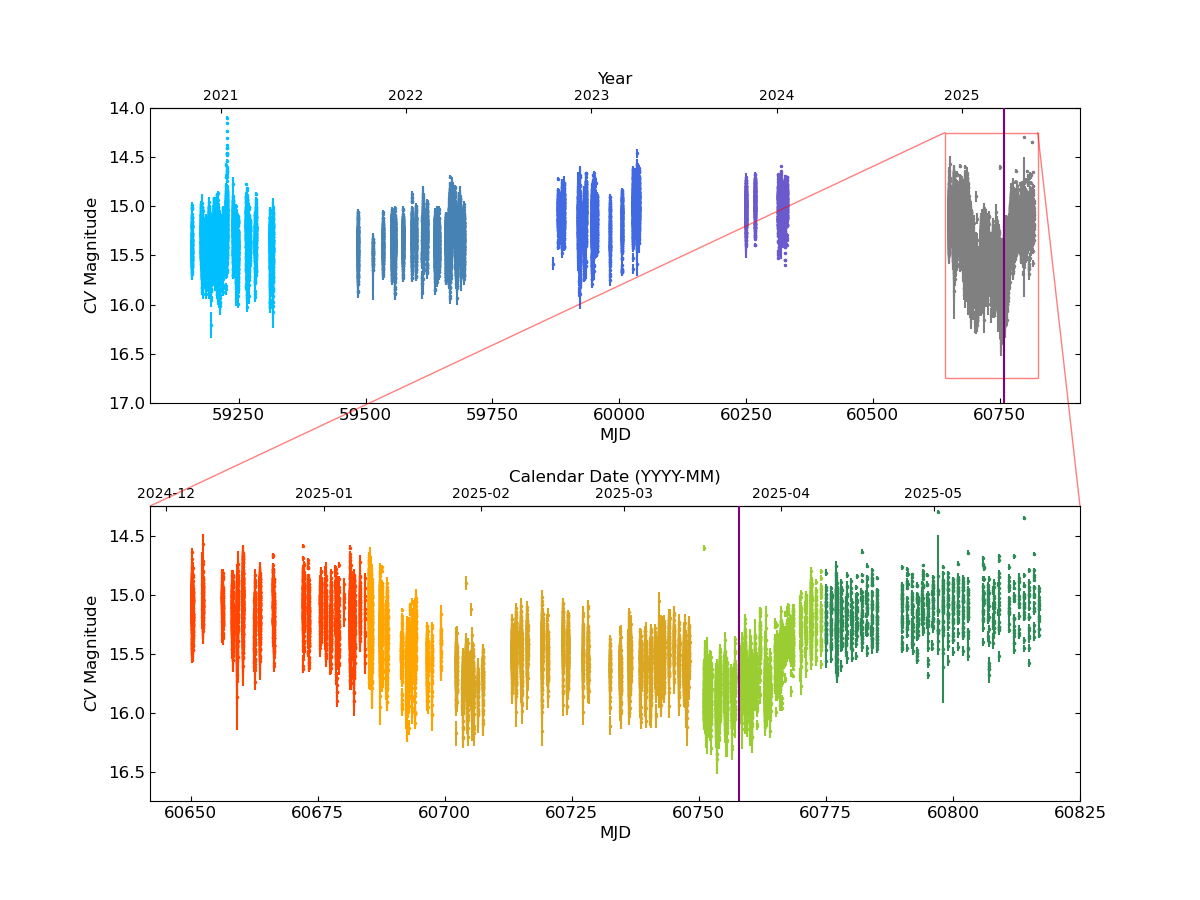}
    \caption{{\em Upper Panel}: Long-term AAVSO light curve of \src\ from 2020 to 2025. Observations were taken in a clear filter and calibrated using a $V$-band zero-point ($CV$-band). {\em Lower Panel}: A zoomed in portion of the AAVSO light curve indicated by the red box in the upper panel and highlighting the low state of the source. Colors represent the epochs that we have broken the light curves down into for timing analysis. Though the first four epochs represent individual observing seasons, the low state is split into multiple epochs. In both panels, the purple line represents the time of the \xmm\ observing campaign.}
    \label{fig:AAVSO_LC}
\end{figure*}

We also utilized public photometry from the American Association of Variable Star Observers (AAVSO) to examine the long-term timing properties of \src. Photometry dating back to June 2020 was downloaded from the AAVSO database and was filtered such that only $CV$-band (i.e. unfiltered data with a $V$-band zero-point) was considered as this was the most prevalent data in the archive. We split the data into epochs typically determined by the annual visibility period of \src\ from the ground. We also filtered the photometry such that no data overlapped in time in the case of two or more observers observing a source simultaneously, prioritizing the observer who provided the most data in each epoch and then including all non-simultaneous data from observers who observed the source during the same epoch. The AAVSO light curves are presented in Fig. \ref{fig:AAVSO_LC}. We also specifically show the light curve from AAVSO observer J. Hambsch (observer code: HMB) in Fig. \ref{fig:XMMcampaign_LC} as it overlaps with the \xmm\ observation.

All time-series photometry described above has been corrected to the solar system barycenter using Astropy's {\tt time} package.

\subsection{Additional X-ray observations}

\subsubsection{Neil Gehrels Swift Observatory}
\label{sec:swift}

In preparation for the \xmm\ campaign, we also requested Target-of-Opportunity (ToO) observations of \src\ with the {\em Neil Gehrels Swift Observatory} (\swift) to monitor the flux of the source \citep{Shaw-2025}. Though \swift\ observed \src\ five times in 2025 March in the lead up to the \xmm\ observation, for the purposes of this work we focus on the longest of those observations. \src\ was observed by \swift\ on 2025 March 18 at 05:48 UT for 1.9 ks (ObsID: 00088624008). The X-ray Telescope \citep[XRT;][]{Burrows-2005} operated in photon counting mode. Using {\sc xselect}, spectra were extracted from a circular aperture of radius 20 pixels ($\sim40\arcsec$), centered on the source. Background spectra were extracted from a circular aperture of radius 50 pixels ($\sim118\arcsec$), centered on a source-free region. An ancillary response file was generated using the FTOOL {\tt xrtmkarf} (included in {\sc HEAsoft} v6.35.1) and the response matrix file was obtained from the calibration database (CALDB). The 0.3--10 keV \swift/XRT count rate was found to be $0.23$ counts s$^{-1}$

We applied the same spectral extraction methodology to two archival \swift/XRT observations of \src\ that occured when the source was in its typical high state: ObsID 00037144009 (2012 September 06 at 00:20, 7 ks exposure, 0.3--10 keV count rate $0.11$ counts s$^{-1}$) and ObsID 00088624003 (2018 October 12 at 05:30 UT, 4.9 ks exposure, 0.3--10 keV count rate $0.15$ counts s$^{-1}$). For all \swift-XRT observations presented in this work, spectra have been grouped such that each bin has a minimum of 20 counts.


\subsubsection{Suzaku}

In order to be able to draw comparisons between the X-ray properties of \src\ during its low state and those during its `normal' state, we searched the High Energy Astrophysics Science Archive Research Center (HEASARC) for archival X-ray observations of the source. Aside from the \swift\ observations discussed in Section \ref{sec:swift}, we also examined a long 2009 \suz\ observation. On 2009 April 11 at 12:11 UT, \suz\ observed \src\ for 47 ks (ObsID: 404029010; see \citealt{Yuasa-2010}).

In 2009 \suz\ \citep{Mitsuda-2007} had three operating CCDs that formed the X-ray imaging Spectrometer (XIS), along with the Hard X-ray Detector (HXD). As we are drawing comparisons with \xmm, we focus here on the XIS data only. We performed full reprocessing and screening using the FTOOL {\tt Aepipeline} provided by {\sc HEAsoft} v6.35.1. Source spectra and light curves were extracted from a circular aperture of radius $100\arcsec$ centered on the source, and background spectra and light curves were extracted from a circular aperture of radius $150\arcsec$ centered on a source-free region. Response matrices and ancillary responses were generated with the FTOOLS {\tt xisrmfgen} and {\tt xissimarfgen} \citep{Ishisaki-2007}, respectively. Spectra from each of the XIS detectors have been grouped such that each bin has a minimum of 20 counts. Light curves were extracted in the 0.2--12 keV energy range for each active detector, and combined with the tool {\tt lcmath}. The combined XIS 0, 1 and 3 light curve had an average 0.2--12 keV count rate of $0.97$ counts s$^{-1}$.

\section{Analysis and Results}
\label{sec:Analysis}

\subsection{Timing Analysis}
\label{sec:Timing}

For all timing analysis performed in this work, we assume a spin frequency of $\omega=1.095\times10^{-3}$ Hz \citep[$P_{\rm spin}=913.5$ s,][]{Bonnardeau-2016,Patterson-2020} and an orbital frequency of $\Omega=8.589\times10^{-5}$ Hz \citep[$P_{\rm orb}=3.23$ h,][]{Bonnardeau-2016}. We note that the spin period of \src\ is known very precisely \citep[see also][]{Bruch-2025}. However, for the purpose of identifying the nature of various peaks in the periodograms, such high precision isn't required.

\subsubsection{XMM-Newton Campaign}

\begin{figure*}
    \centering
    \includegraphics[width=0.95\textwidth]{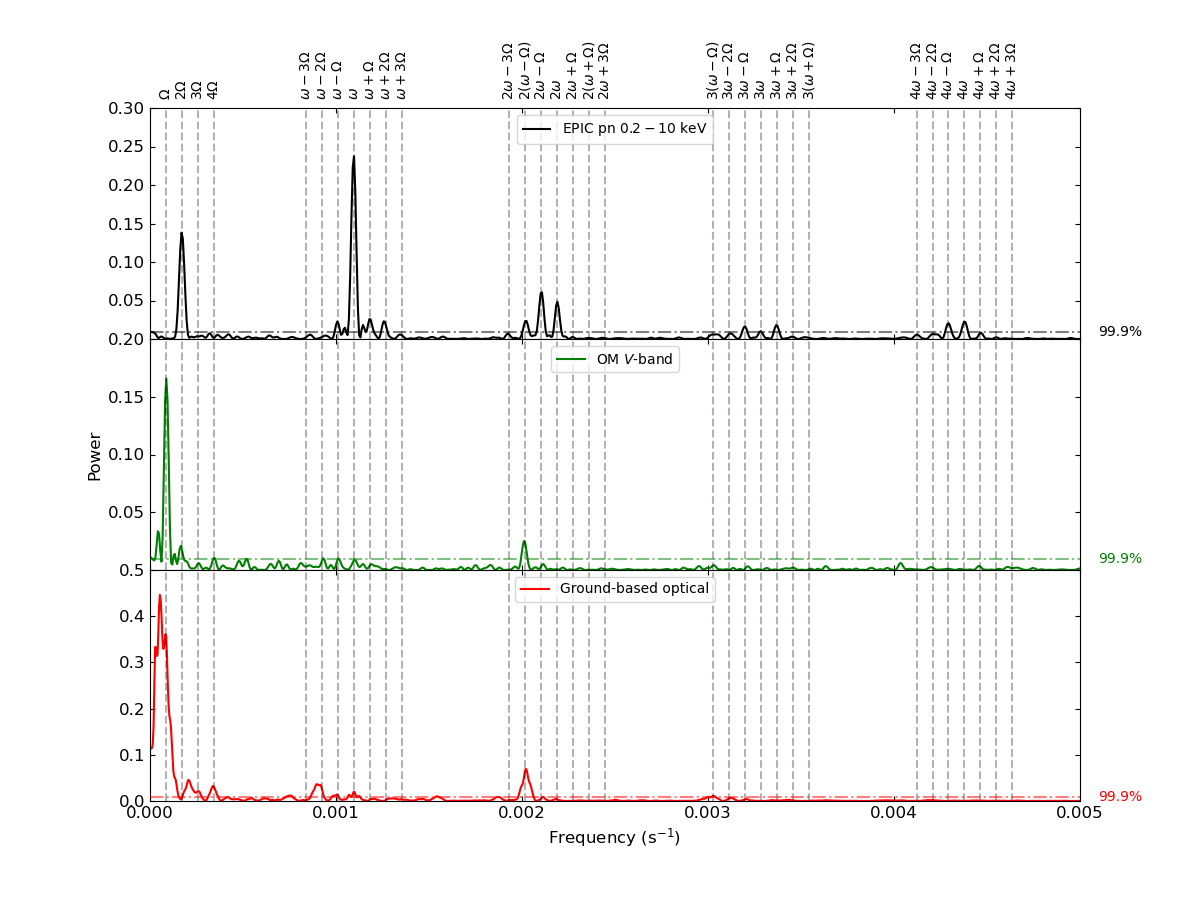}
    \caption{{\em Upper Panel}: Lomb-Scargle periodogram of the \xmm\ EPIC/pn 0.2--10 keV light curve of \src. {\em Middle Panel}: Lomb-Scargle periodogram of the \xmm\ EPIC/OM light curve of \src. {\em Lower Panel}: Lomb-Scargle periodogram of the non-overlapping ground-based optical light curve of \src. In all panels, vertical dashed lines represent known periodicities in the system, including the orbital and spin frequencies ($\Omega$ and $\omega$, respectively) and their multiples, as well as beat and sideband frequencies. $99.9\%$ significance thresholds for the Lomb-Scargle power are shown as horizontal dot-dashed lines in each panel.}
    \label{fig:XMM_LS}
\end{figure*}

To investigate the timing properties of \src, we generated Lomb-Scargle periodograms \citep{Lomb-1976,Scargle-1982} using the {\tt LombScargle} class included in Astropy. The periodograms of the \xmm\ light curves are shown in Fig.\  \ref{fig:XMM_LS}, along with a periodogram calculated from the simultaneous ground-based optical light curves. For the ground-based photometry we only used non-overlapping data, and normalized the light curves in order to reduce the chance of spurious signals appearing in the periodogram. For each periodogram, we calculated a $99.9\%$ significance threshold for the power by randomly shuffling the light curve fluxes (or magnitudes) while keeping the time stamps the same, effectively creating a randomized light curve with the same sampling as the original data. For each panel of Fig.\  \ref{fig:XMM_LS} we calculated the peak power for 10,000 such randomized light curves, from which we derived the $99.9\%$ significance threshold. 

It is immediately clear from Fig.\  \ref{fig:XMM_LS} that the timing properties of \src\ during the low state are wavelength-dependent. At X-ray energies, the strongest peaks in the periodogram are at the spin frequency $\omega$ (and $2\omega$) and twice the orbital frequency $2\Omega$. However, there is significant power at numerous sideband frequencies, with the strongest peak at $2\omega-\Omega$. However, at optical wavelengths, the spin, while still significant at the $99.9\%$ level in at least the ground-based light curves, is very weak. Instead, the strongest (non-orbital) timing signature at optical wavelengths is twice the beat frequency, $2(\omega-\Omega)$.

\subsubsection{AAVSO}

\begin{figure*}
    \centering
    \includegraphics[width=0.95\textwidth]{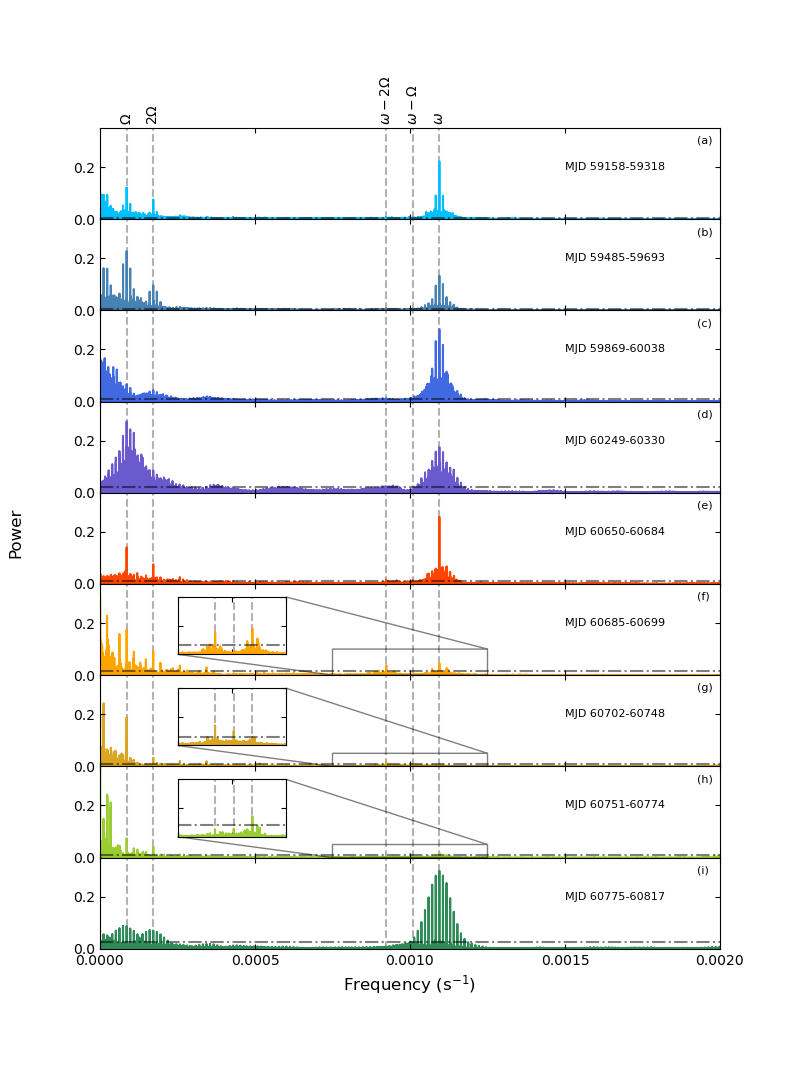}
    \caption{Lomb-Scargle periodograms of the AAVSO light curves of \src. The colors match the epochs defined in Fig.\  \ref{fig:AAVSO_LC}, and the relevant MJD ranges are labeled in each panel. The vertical dashed lines represent known periodicities related to the system and the dot-dashed horizontal lines in each panel indicated the $99.9\%$ confidence level. }
    \label{fig:AAVSO_LS}
\end{figure*}

The periodograms in Fig.\ \ref{fig:XMM_LS} are just a snapshot of the temporal properties of \src\ during its low state. In order to investigate the possibility of changing accretion modes, we must compare the timing properties of \src\ in the low state with those out of it. At optical wavelengths, we can do this by calculating a series of periodograms using the AAVSO light curves presented in Fig.\  \ref{fig:AAVSO_LC}. The periodograms are shown in Fig.\  \ref{fig:AAVSO_LS}. In panels (a) -- (e) we see that the light curves are dominated by variability at $\Omega$ and $\omega$. However, as the optical flux of \src\ starts to decline, so does the strength of the peak at $\omega$, to the point where it is below the $99.9\%$ significance line in panel (g), as noted by \citet{Littlefield-2025}. During the decline and at the base of the low state, we see significant power at $\omega-2\Omega$ and, in panel (g) only, $\omega-\Omega$. As the flux recovers to levels comparable with those prior to the low state, so does the 
power at $\omega$.

\subsubsection{Suzaku}

At X-ray energies, we can draw comparisons between the temporal properties of \src\ during the low state and the `normal' state by performing Lomb-Scargle analysis on the 0.2--12 keV \suz\ light curve of \src\ from 2009. As this is the first reported low state of \src, and there has been continuous optical monitoring of the source annually for many decades \citep{Bonnardeau-2016,Patterson-2020} with no mention of a low state in the literature, it is a reasonable assumption that the source was in its typical high state at the time of the \suz\ observation.

\begin{figure*}[h!]
    \centering
    \includegraphics[width=0.95\textwidth]{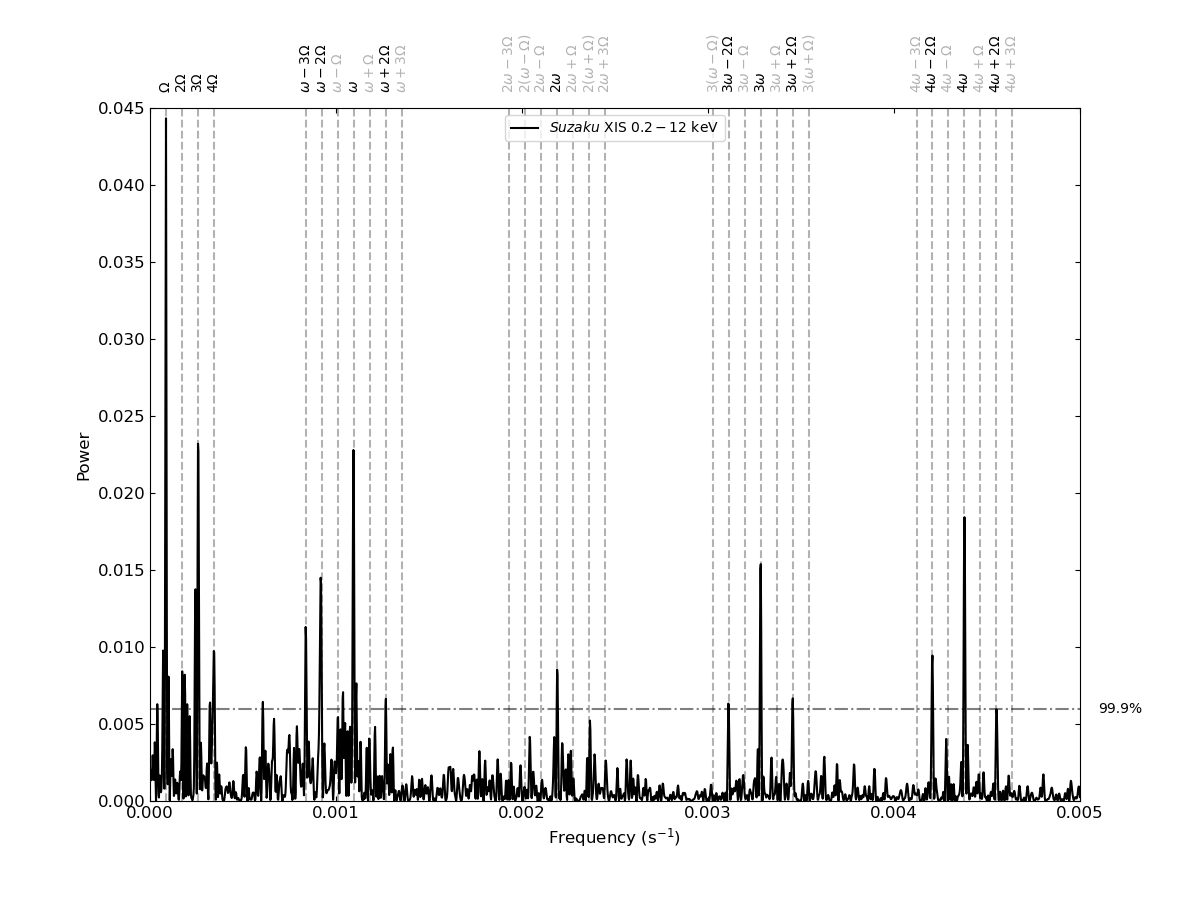}
    \caption{Lomb-Scargle periodogram of the 2009 Apr 11 \suz\ 0.2--12 keV light curves of \src. The vertical dashed lines represent known periodicities related to the system, with undetected periods shown in grey at the top of the figure. The dot-dashed horizontal line represents the $99.9\%$ confidence level.}
    \label{fig:Suzaku_LS}
\end{figure*}

The \suz\ periodogram is shown in Fig.\ \ref{fig:Suzaku_LS}. As in Fig.\  \ref{fig:XMM_LS}, variability at $\omega$ (and its harmonics) is highly significant in the X-ray band. In addition, the \suz\ periodogram exhibits significant power at $\Omega$ and its harmonics, though the strongest peak is at $\Omega$, rather than at $2\Omega$ seen in the \xmm\ periodogram. There are also many $\pm2\Omega$ spin-orbital sidebands apparently present in the \suz\ periodogram. However, we must note that the orbital period of the \suz\ satellite ($P_{{\rm orb}, Suz}=96$ min) is coincidentally very close to half the orbital period of \src, such that this may contaminate the Lomb-Scargle analysis. Despite this, a key difference between the \xmm\ EPIC/pn and \suz\ periodograms is the presence of significant variability at $2\omega-\Omega$ and $2(\omega-\Omega)$ during the low state (Fig. \ref{fig:XMM_LS}; upper panel), which is not present in the \suz\ periodogram. In addition, the \xmm\ EPIC/pn periodogram also shows statistically significant power at the $\pm\Omega$ spin-orbital sidebands that is not present in the archival \suz\ observation.

\subsection{X-ray Spectral Analysis}

\begin{figure}
    \centering
    \includegraphics[width=0.45\textwidth]{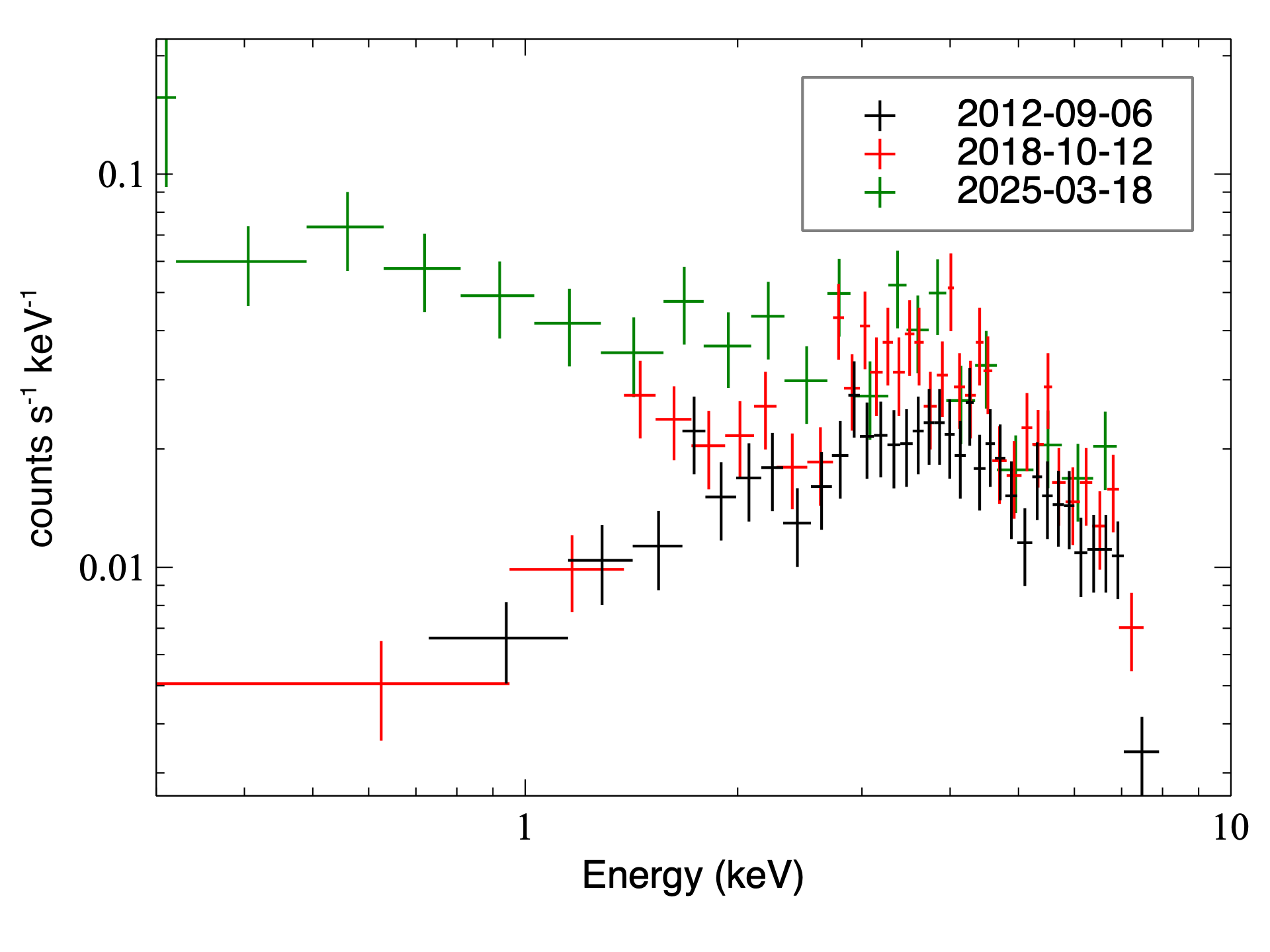}
    \caption{{\em Swift}/XRT spectra of \src\ taken during its regular high state (2012 and 2018; black and red, respectively) and the low state (2025; green). No models have been fit to the spectra.}
    \label{fig:Swift_comp}
\end{figure}

In Fig. \ref{fig:Swift_comp} we show three 0.5--10 keV {\em Swift}/XRT snapshots of \src\ from 2012, 2018 and 2025, with the 2025 observation having occurred during the low state. Even a simplistic comparison such as this highlights a drastic change in the X-ray properties of \src\ during the low state, with a large increase in soft flux. To investigate this further, we must model the X-ray spectra of \src\ both in and out of the low state.

Here, we model the X-ray spectra from the typical high state of \src\ (\suz) and from the low state (\xmm). All modeling is performed in {\sc xspec} v12.15.0 using \citet{Wilms-2000} abundances, \citet{Verner-1996} photoionization cross-sections and the $\chi^2$ fitting statistic.

\subsubsection{The Suzaku high state spectrum}

\citet{Ezuka-1999} present a 0.5--10 keV Advanced Satellite for Cosmology and Astrophysics Solid-state Imaging Spectrometer and Gas Imaging Spectrometer \citep[{\em ASCA}/SIS and GIS][]{Tanaka-1994} spectrum of \src\ as part of a survey of 23 mCVs. \citet{Ezuka-1999} found that the 0.5--10 keV spectrum could be well-described by a thermal bremsstrahlung model modified by three partial covering absorbers, in addition to three Gaussians to represent the iron line complex in the 6--7 keV range. This is a common way to model the X-ray spectra of IPs, though sometimes one or two single-temperature thermal plasmas are used in place of the thermal bremsstrahlung component \citep[see e.g.][]{Mukai-2015,Kennedy-2017}.

We fit the model of \citet{Ezuka-1999} to the archival \suz/XIS spectra of \src\ in the 0.8--10 keV energy range. We ignore photons with energies 1.7--1.9 keV and 2.2--2.4 keV due to calibration issues in those energy ranges. The full model, as written in {\sc xspec} notation is {\tt constant*pcfabs*pcfabs*pcfabs*(bremss+gaussian)}, where {\tt constant} represents a cross-normalization constant between the three XIS detectors. We note that \citet{Ezuka-1999} do not include an interstellar absorption component such as {\tt tbabs} \citep{Wilms-2000}. We find an acceptable fit, with $\chi^2_{\nu}=1.13$\footnote{Reduced chi-squared statistic: $\chi^2$ per degrees of freedom (dof)} ($\chi^2/{\rm dof}=1897/1682$). However, we find that one of the partial covering components is poorly constrained, with an unbound $N_{\rm H}$ and covering fraction. Thus we consider it an unnecessary inclusion in the model.


We then model the \suz\ spectrum of \src\ with an isobaric cooling flow model \citep{Mushotzky-1988,Mukai-2003}, modified by two partial covering components and an interstellar absorption component, which we freeze to the Galactic value of $N_{\rm H}=0.06\times10^{22}$ cm$^{-2}$ \citep{HI4PI-2016}. The cooling flow model ({\tt mkcflow} in {\sc xspec}) has been used to successfully describe the X-ray spectra of many IPs \citep[see e.g.][]{Mukai-2015} and is a much more accurate physical representation of the post-shock flow above the surface of the WD than a single-temperature bremsstrahlung model. (Even more detailed post-shock models have been constructed to fit the full spectra of IPs, extending up to $\sim$80 keV, e.g. \citealt{Suleimanov-2005,Yuasa-2010,Suleimanov-2019}. As we do not use data above 10 keV or attempt to constrain the WD mass, the current model is sufficient.) In {\tt xspec} notation, our model is written as {\tt constant*tbabs*pcfabs*pcfabs*(mkcflow + gaussian)}, where the additional Gaussian component represents the neutral Fe K$\alpha$ line. We find that this model describes the high-state spectrum well ($\chi^2_{\nu}=1.12$; 1683 dof) and we list the best-fit parameters in Table \ref{tab:fits}. We choose two partial covering components to represent two distinct components of the intrinsic absorber.  \src\ shows grazing eclipses in the optical (see Fig. \ref{fig:XMMcampaign_LC}), and thus is a moderately inclined binary \citep[$i\sim55^{\circ}-75^{\circ}$;][]{deMartino-1995}. Therefore, we might expect some of the intrinsic absorption to come from the disk, similar to FO\,Aqr \citep{Evans-2004,Kennedy-2017}, while some will come from the pre-shock flow, as for the majority of IPs \citep[see e.g.][]{Mukai-2017}. We call the model with two partial covering components `Model 1.'

\citet{Done-1998} note that multiple discrete partial covering models, each with their own hydrogen column densities ($N_{\rm H}$) and covering fractions ($f$), are a poor approximation for the complex absorption of X-ray photons by the pre-shock flow. A distribution of covering fractions as a function of $N_{\rm H}$, approximated as a power law, provides a much more physical representation of the absorption. We therefore replace one of the partial covering components with an ionized complex absorber {\tt zxipab} \citep{Islam-2021}, a modification to the \citet{Done-1998} model. We call this model `Model 2' and it is written as {\tt constant*tbabs*pcfabs*zxipab*(mkcflow + gaussian)} in {\tt xspec} notation. We find that Model 2 results in a marginally better fit than Model 1 ($\Delta\chi^2=-6$; for one fewer dof). However, they are almost indistinguishable. We therefore show the best fit parameters for both models in Table \ref{tab:fits}.

We find hydrogen column densities of $N_{\rm H,pc1}=3.2\pm0.4\times10^{22}$ cm$^{-2}$ and $N_{\rm H,pc2}=14.2^{+3.1}_{-1.9}\times10^{22}$ cm$^{-2}$ with covering fractions of $f_{pc1}=0.93\pm0.01$ and $f_{pc2}=0.67\pm0.04$ for the first and second partial covering components in Model 1, respectively, indicating significant intrinsic absorption. In Model 2, we find $N_{\rm H,pc1}=3.2^{+1.2}_{-0.5}\times10^{22}$ cm$^{-2}$ and $f_{pc1}=0.78^{+0.06}_{-0.15}$ for the lone partial covering component. For the complex absorber in Model 2, we find a maximum column density $N_{\rm H,max}=20.2^{+9.3}_{-4.6}\times10^{22}$ cm$^{-2}$ with a power-law index of the covering fraction $\beta=-0.38^{+0.18}_{-0.12}$ \citep[where $f\propto N_{\rm H}^\beta$ and $\int^{N_{\rm H,max}}_{N_{\rm H,min}}f=1.0$;][]{Islam-2021}.\footnote{We freeze the minimum column density to $N_{\rm H,min}=10^{15}$ cm$^{-2}$, essentially zero.} A negative value of $\beta$ is indicative of a decreasing value of $f$ with increasing $N_{\rm H}$ and, in the specific case of the fit to the \suz\ spectrum of \src\ in the high state, $f=0$ at $N_{\rm H}=20.2^{+9.3}_{-4.6}\times10^{22}$ cm$^{-2}$.

\begin{figure}
    \centering
    \includegraphics[width=0.45\textwidth]{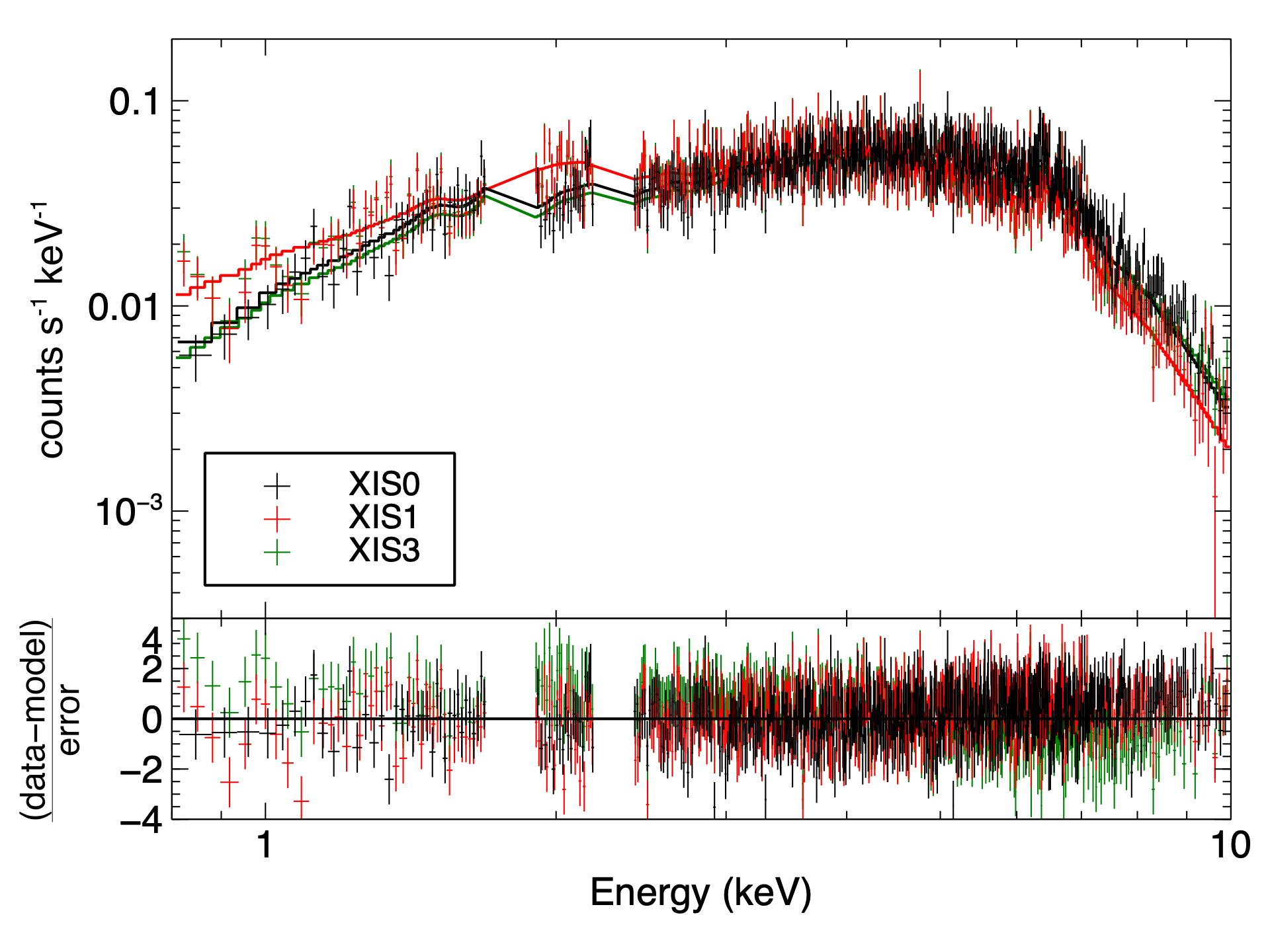}
    \caption{Archival \suz/XIS spectra of \src\ during the high state. The spectra have been fit with a cooling flow model modified by partial covering and a complex absorber, denoted Model 2 in the text. Residuals are shown in the lower panel.}
    \label{fig:suz_spec}
\end{figure}

We find an absorbed flux of $\approx1.8\times10^{-11}$ erg cm$^{-2}$ s$^{-1}$ in the 0.5--10 keV energy range, with almost all of that flux emerging in the 2--10 keV band ($\approx1.8\times10^{-11}$). We find an unabsorbed flux of $\approx3.7\times10^{-11}$ erg cm$^{-2}$ s$^{-1}$ and $\approx2.7\times10^{-11}$ erg cm$^{-2}$ s$^{-1}$ in the 0.5--10 keV and 2--10 keV bands, respectively. The quoted fluxes are for Model 1, but we find consistent results for Model 2. We plot the \suz\ spectra and the Model 2 fit in Fig.\ \ref{fig:suz_spec}.

\begin{table*}[]
\caption{Best-fit parameters of two model fits to the X-ray spectra of \src\ from two epochs. The \suz\ observations occurred when the source was in its typical high state, while the \xmm\ observations took place during the 2025 low state. Models are described in the text, but we incorporate an additional Gaussian (G2) when modeling the \xmm\ spectra to represent unresolved emission lines below 1 keV.}
\centering
\begin{tabular}{lcccc}
\hline
          & \multicolumn{2}{c}{\suz}       & \multicolumn{2}{c}{\xmm}  \\
& \multicolumn{2}{c}{(high state)} & \multicolumn{2}{c}{(low state)} \\
\hline
Parameter & Model 1 & Model 2 & Model 1 & Model 2 \\
\hline
\hline
     $N_{\rm H}$ ($\times 10^{22}~ {\rm cm}^{-2}$) &$0.06^*$&$0.06^*$&$0.06^*$&$0.06^*$\\
     $N_{\rm H,pc1}$ ($\times 10^{22}~ {\rm cm}^{-2}$) &$3.2\pm0.4$&$3.2^{+1.2}_{-0.5}$&$2.9\pm0.4$&$7.3^{+2.1}_{-1.4}$\\
     $f_{\rm pc1}$ &$0.93\pm0.01$&$0.78^{+0.06}_{-0.15}$&$0.63^{+0.02}_{-0.04}$&$0.30^{+0.07}_{-0.09}$\\
     $N_{\rm H,pc2}$ ($\times 10^{22}~ {\rm cm}^{-2}$) &$14.2^{+3.1}_{-1.9}$&\nodata &$28.5^{+5.8}_{-5.1}$&\nodata \\
     $f_{\rm pc2}$ &$0.67\pm0.04$&\nodata&$0.55\pm0.04$&\nodata\\
     $N_{\rm H,max}$ ($\times 10^{22}~ {\rm cm}^{-2}$) &\nodata&$20.2^{+9.3}_{-4.6}$ &\nodata&$>64.7$ \\
     $\beta$ &\nodata&$-0.38^{+0.18}_{-0.12}$&\nodata&$-0.68\pm0.04$\\
     $\log_{10}\xi$ &\nodata&$0.3^{+0.6}_{-0.7}$&\nodata&$0.6\pm0.2$\\
     $T_{\rm max}$ (keV) & $>42.7$ & $20.2^{+9.3}_{-4.6}$ & $34.1^{+17.5}_{-6.6}$ & $24.3^{+9.9}_{-3.3}$ \\
     abundance (rel. to solar) & $0.49^{+0.17}_{-0.16}$ & $0.28^{+0.22}_{-0.10}$& $0.35^{+0.15}_{-0.11}$ & $0.25^{+0.07}_{-0.06}$\\
     Normalization ({\tt mkcflow}; $\times 10^{-10}~M_{\odot}~{\rm yr}^{-1}$) & $4.7^{+2.4}_{-0.8}$ & $7.6^{+3.7}_{-4.2}$& $9.5^{+2.4}_{-3.1}$ & $16.5^{+22.5}_{-6.1}$\\
     $E_{\rm G1}$ (keV) & $6.40^{+0.04}_{-0.03}$ & $6.40\pm0.03$& $6.40^{+0.04}_{-0.03}$ & $6.40\pm0.03$\\
     $\sigma_{\rm G1}$ (keV) & $0.12^{+0.06}_{-0.05}$ & $0.09\pm0.05$& $0.09\pm0.06$ & $<0.11$\\
     Normalization (G1; $\times 10^{-5}$ photon cm$^{-2}$ s$^{-1}$) & $3.7^{+0.9}_{-0.8}$ & $3.0^{+0.8}_{-0.7}$ & $4.0^{+1.2}_{-1.0}$ & $4.0^{+4.4}_{-1.1}$\\
     $E_{\rm G2}$ (keV) & \nodata  & \nodata & $0.54^{+0.01}_{-0.02}$ & $0.52^{+0.03}_{-0.05}$\\
     $\sigma_{\rm G2}$ (keV) & \nodata  & \nodata & $0.08\pm0.01$ & $0.10^{+0.03}_{-0.02}$\\
     Normalization (G2; $\times 10^{-3}$ photon cm$^{-2}$ s$^{-1}$) & \nodata & \nodata & $4.5^{+1.0}_{-0.9}$ & $8.8^{+15.3}_{-2.8}$\\
     \hline
     $C_{\rm XIS1}$& $1.09\pm0.02$ & $1.09\pm0.02$ & \nodata & \nodata \\
     $C_{\rm XIS3}$& $0.82\pm0.02$ & $0.82\pm0.02$ & \nodata & \nodata \\
     $C_{\rm pn}$& \nodata & \nodata & $0.67\pm0.01$ & $0.67\pm0.01$ \\
     $C_{\rm MOS1}$& \nodata & \nodata & $0.98\pm0.02$ & $0.98\pm0.02$ \\
     \hline
     $F_{\rm abs, 0.5-10}$ ($\times10^{-11}$ erg cm$^{-2}$ s$^{-1}$) &$1.8\pm0.3$& $1.8^\dagger$&$2.3^{+1.1}_{-1.2}$&$2.3^\dagger$\\
     $F_{\rm unabs, 0.5-10}$ ($\times10^{-11}$ erg cm$^{-2}$ s$^{-1}$) &$3.7\pm0.1$& $3.9\pm0.3$&$5.0\pm0.3$&$6.9^{+6.0}_{-1.1}$\\
     $F_{\rm abs, 2-10}$ ($\times10^{-11}$ erg cm$^{-2}$ s$^{-1}$) &$1.8\pm0.2$& $1.8^\dagger$&$2.0^{+1.3}_{-1.0}$&$2.0^\dagger$\\
     $F_{\rm unabs, 2-10}$ ($\times10^{-11}$ erg cm$^{-2}$ s$^{-1}$) &$2.7\pm0.1$& $2.6\pm0.2$&$3.2^{+0.2}_{-0.1}$&$4.0^{+3.7}_{-0.6}$\\
     \hline
     $\chi^2$/dof & 1886/1683 & 1880/1682 & 1940/1894 & 1929/1893\\
     
\hline
\end{tabular}
\\
\raggedright$^*$ Frozen\\
\raggedright$^\dagger$ Error analysis did not converge\\
$N_{\rm H}$: Hydrogen column density\\
$N_{\rm H,pc1/2}$: Hydrogen column density for partial covering component 1/2\\
$f_{\rm pc1/2}$: Covering fraction for partial covering component 1/2\\
$N_{\rm H,max}$: Maximum hydrogen column density for the {\tt zxipab} model component\\
$\beta$: Covering fraction power-law index ($f=N_{\rm H}^{\beta}$)\\
$\log_{10}\xi$: Ionization parameter\\
$T_{\rm max}$: Maximum temperature of the cooling flow\\
$E_{\rm G1/2}$ Central energy of Gaussian 1/2\\
$\sigma_{\rm G1/2}$ Width of Gaussian 1/2\\
$C$: Cross-normalization constant (relative to XIS2 for \suz\ and MOS1 for \xmm)\\
$F_{\rm abs/unabs}$: Absorbed/unabsorbed flux in the indicated energy band
\label{tab:fits}
\end{table*}

\subsubsection{The XMM-Newton low state spectrum}
\label{sec:XMMspec}

\begin{figure}
    \centering
    \includegraphics[width=0.45\textwidth]{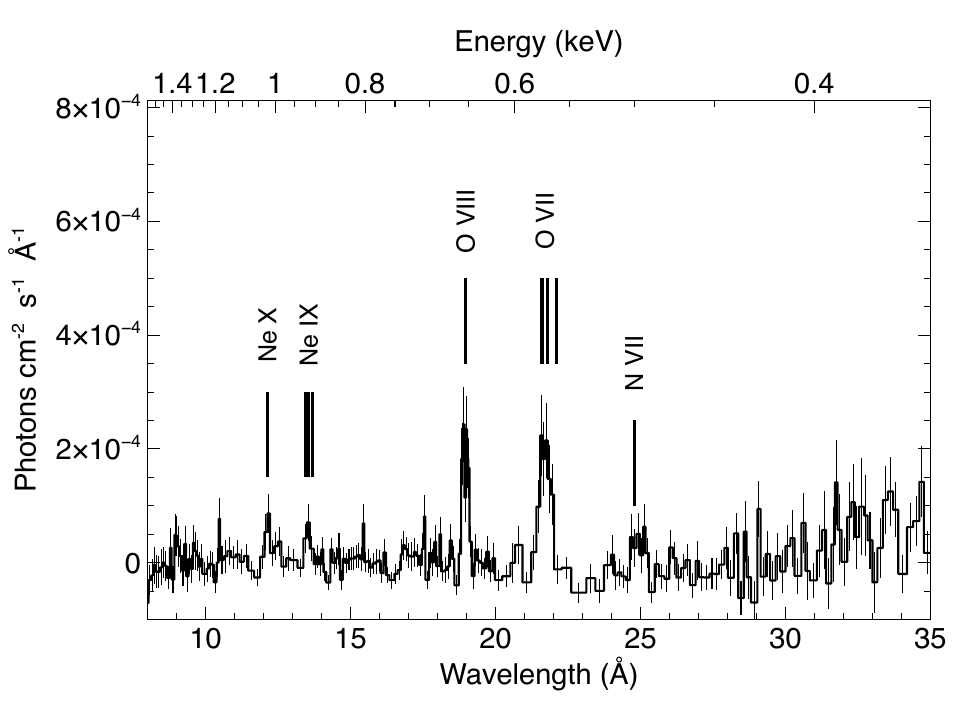}
    \caption{Residuals of a phenomenological fit to the \xmm/RGS spectrum of \src, with some spectral features labeled.}
    \label{fig:RGS}
\end{figure}

We model the 0.5--10 keV low-state \xmm/EPIC spectra of \src\ in a similar manner to the \suz\ spectra, using Model 1 and Model 2. However, the initial fit reveals a low-energy excess below $\lesssim0.8$ keV. To investigate this excess further we reduce and combine the data from the two Reflection Grating Spectrometer \citep[RGS;]{den_Herder-2001} detectors onboard \xmm. We used the {\tt rgsproc} and {\tt rgscombine} tasks in {\sc sas} and fit a phenomenological model of an absorbed power law (with partial covering; {\tt tbabs*pcfabs*powerlaw}) to the resultant spectrum and plot the residuals in Fig.\ \ref{fig:RGS}. Though we do not make any attempt to statistically identify lines \citep[for example through a blind search, e.g.][]{Shaw-2022}, it is clear that there are narrow lines present, with the strongest being coincident with the rest wavelengths of O {\sc viii} Ly$\alpha$ and the  O {\sc vii} He-like triplet. 

We therefore include an additional Gaussian component in our models when fitting the \xmm/EPIC spectra of \src, initialized at a line energy of 0.6 keV, to account for the unresolved oxygen lines. The updated Model 1 is thus {\tt constant*tbabs*pcfabs*pcfabs*(mkcflow + gaussian + gaussian)}, while Model 2 becomes {\tt constant*tbabs*pcfabs*zxipab*(mkcflow + gaussian + gaussian)}. We find that both models describe the data very well, with fit statistics of $\chi^2_{\nu}=1.02$ (1894 dof) and $\chi^2_{\nu}=1.02$ (1893 dof) for Model 1 and 2, respectively. Similar to the \suz\ model fits, Model 2 results in a marginally better fit than Model 1 ($\Delta\chi^2=-11$; for one fewer dof) but we must note that $N_{\rm H,max}$ is unconstrained in the best-fit Model 2. We also remind the reader that the cross-normalization constant to EPIC-pn, $C_{\rm pn}$, is significantly lower than unity, owing to our piecewise pile-up correction method detailed in Section \ref{sec:intro}. The spectral shape in EPIC-pn is consistent with that measured by the MOS detectors.

For Model 1, we find $N_{\rm H,pc1}=2.9\pm0.4\times10^{22}$ cm$^{-2}$ and $N_{\rm H,pc2}=28.5^{+5.8}_{-5.1}\times10^{22}$ cm$^{-2}$ with covering fractions of $f_{pc1}=0.63^{+0.02}_{-0.04}$ and $f_{pc2}=0.55\pm0.04$, respectively. The covering fractions are significantly smaller than the best-fit values for the \suz\ spectrum, indicating a drastic reduction in absorption local to the source. For Model 2, we find $N_{\rm H,pc1}=7.3^{+2.1}_{-1.4}\times10^{22}$ cm$^{-2}$ and $f_{pc1}=0.30^{+0.07}_{-0.09}$ for the lone partial covering component, once again indicating a drastic reduction in local absorption. For the complex absorber, we find a maximum column density $N_{\rm H,max}>64.7\times10^{22}$ cm$^{-2}$ and $\beta=-0.68\pm0.04$, which is significantly steeper (and for a higher $N_{\rm H,max}$) than in the high state.

\begin{figure}
    \centering
    \includegraphics[width=0.45\textwidth]{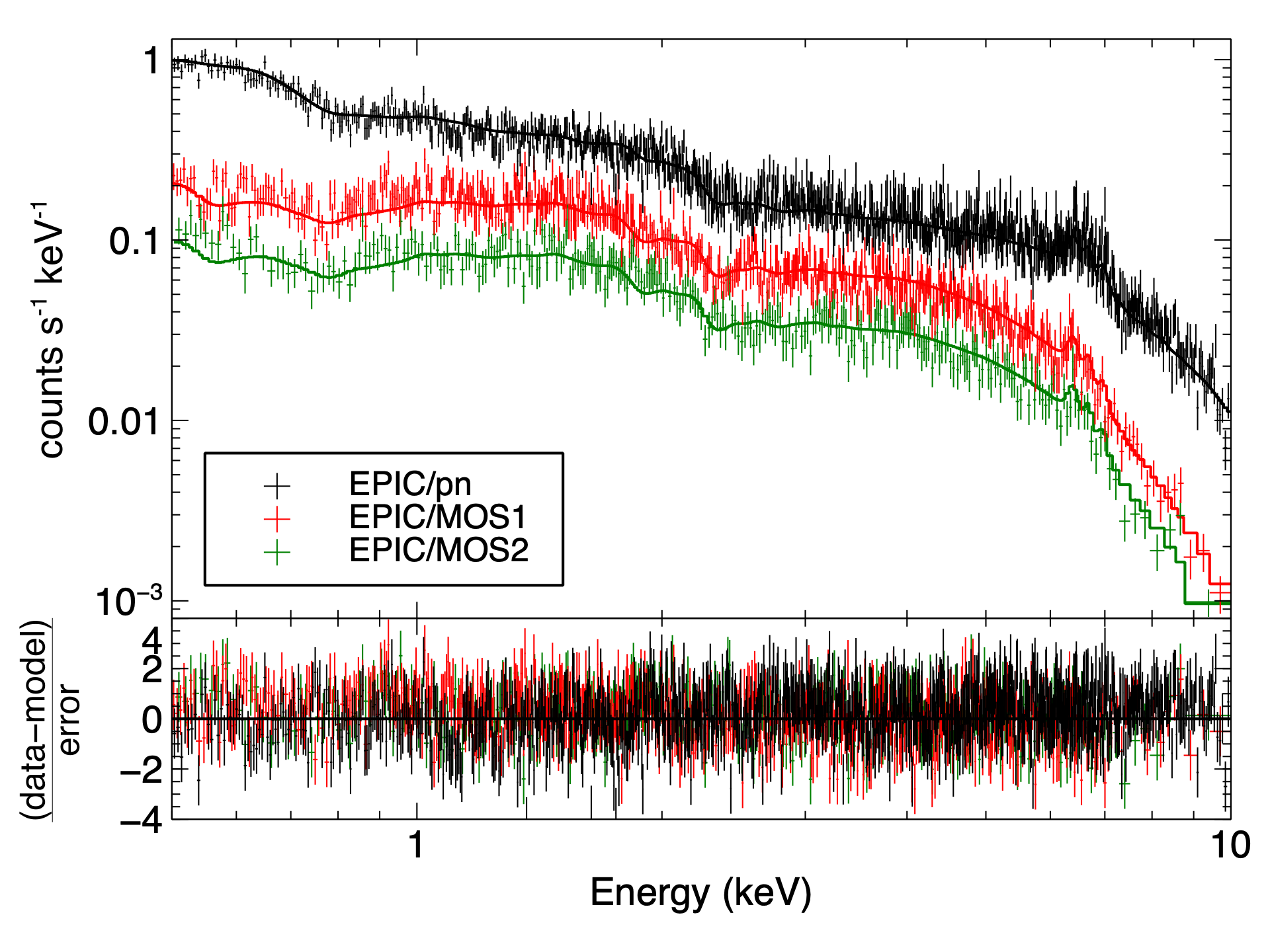}
    \caption{\xmm/EPIC spectra of \src\ during the low state in March 2025. The spectra have been fit with Model 2. Residuals are shown in the lower panel.}
    \label{fig:xmm_spec}
\end{figure}

We find an absorbed flux of 
$2.3\times10^{-11}$ erg cm$^{-2}$ s$^{-1}$ in the 0.5--10 keV energy range and 
$2.0\times10^{-11}$ in the 2--10 keV band. We find an unabsorbed flux of 
$5.0\times10^{-11}$ erg cm$^{-2}$ s$^{-1}$ and 
$3.2\times10^{-11}$ erg cm$^{-2}$ s$^{-1}$ in the 0.5--10 keV and 2--10 keV bands, respectively. The quoted fluxes are for Model 1, but we find the unabsorbed fluxes in Model 2 have large uncertainties due to the unconstrained value of $N_{\rm H,max}$. Regardless, we find that the X-ray flux of \src\ has not decreased alongside the optical flux during the low state when compared with the archival \suz\ data and, in fact,  appears to have increased overall. As Fig.\ \ref{fig:Swift_comp} suggests, there is a significant increase in soft X-ray flux during the low state, coincident with the decrease in intrinsic absorption. We show the \xmm\ spectra, fit with Model 2, in Fig.\ \ref{fig:xmm_spec}.

\section{Discussion}
\label{sec:discussion}

Although low states have been studied in numerous IPs \citep[e.g.][]{Garnavich-1988,Kennedy-2017,Littlefield-2020,Covington-2022}, this marks only the second time that deep X-ray observations of an IP low state have been reported \citep[after FO Aqr;][]{Kennedy-2017}. X-rays allow us to study the post-shock region close to the surface of the white dwarf, whereas optical photons mostly originate in the accretion disk. Simultaneous optical and X-ray observations allow us to develop a more complete picture of the accretion flow during a low state. Moreover, \src\ is well-studied enough that we can make meaningful comparisons with observations made during its typical high state.

\subsection{Optical and X-ray timing}
\label{sec:discussion:timing}

Fig.\  \ref{fig:AAVSO_LC} highlights the dense optical photometric coverage of \src\ by observers of the AAVSO, and we show the evolution of the timing properties in the optical band in Fig.\  \ref{fig:AAVSO_LS}. The X-ray coverage, by comparison, is much more sparse. However, we are able to draw comparisons between the X-ray timing properties in the high state and the low state by examining an archival \suz\ observation of \src.

In Fig.\ \ref{fig:AAVSO_LS} we see that the optical light curves of \src\ are typically dominated on short (sub-h) timescales by variability at the spin frequency of the WD, $\omega$. However, as the optical flux starts to decline, so does the strength of the variability. In panel (g), when the optical light curve (Fig.\ \ref{fig:AAVSO_LC}) appears to have bottomed out, the spin variability is no longer significant at the 99.9\% level and instead the strongest short term variability, though still weak, is at $\omega-2\Omega$. This change in the dominant frequencies in the optical periodogram is indicative of a change in the accretion mode of the IP. A typical, disk-fed IP will show strong variability on the spin period of the WD due to matter being forced onto field lines at a nearly uniform rate from a Keplerian disk whose inner radius co-rotates with the WD. If, instead, the matter flows from a fixed point in the binary rest frame (stream-fed accretion), the optical variability will appear at an orbital sideband.
This change away from a strong signal at $\omega$ is often associated with low states \citep[e.g.][]{Littlefield-2020,Covington-2022}. 

In Fig.\ \ref{fig:XMM_LS} we show the multi-wavelength timing properties of \src\ during the low state. In the lower two panels we confirm the near-disappearance of the spin variability at optical wavelengths that was seen in the AAVSO periodograms. We also see the emergence of variability on timescales consistent with the $\omega-2\Omega$ and $2(\omega-\Omega$) frequencies. In the upper panel we see that, while the spin of the WD dominates the X-ray light curves, there is significant power at $\omega-\Omega$ and $2\omega-\Omega$ that is not present in the archival \suz\ light curves from the high state (Fig.\ \ref{fig:Suzaku_LS}). Much like with the AAVSO periodograms (Fig. \ref{fig:AAVSO_LS}), this is strongly indicative of a change in accretion mode.

Interestingly, the top panel of Fig.\ \ref{fig:XMM_LS} shows a strong peak at $2\Omega$, while the optical periodograms in the middle and lower panels show variability at $\Omega$, consistent with the orbital variability of the binary. When comparing the \xmm\ EPIC-pn periodogram with that derived from archival \suz\ light curves in Fig.\ \ref{fig:Suzaku_LS} it is clear that the dominant orbital variability has switched from $\Omega$ to $2\Omega$ at X-ray energies.\footnote{We also note that $\Omega$ dominates in the 0.7--10 keV {\em ASCA} light curves \citep{Parker-2005}} This is another indication that the accretion mode has changed.

According to the \citet{Ferrario-1999} model, a switch to a stream-only accretion mode of a moderate-highly inclined source like \src\ should result in the emergence of strong power at sidebands such as $2\omega-\Omega$ at X-ray energies.\footnote{We note that \citet{Ferrario-1999} mislabel this as $\omega-2\Omega$ in their Fig.\ 5.} Simultaneously, the model predicts that, at optical wavelengths, strong power emerges at $\omega-2\Omega$ and $2(\omega-\Omega)$. The emergence of power at these frequencies in the low-state periodograms in Figs.\ \ref{fig:XMM_LS} and \ref{fig:AAVSO_LS}f,g suggests that the accretion stream had become more important during the low state. The near complete disappearance of $\omega$ in the optical periodograms during the low state (Figs.\ \ref{fig:XMM_LS} and \ref{fig:AAVSO_LS}g) may even suggest that the disk had almost completely dissipated, but this cannot be confirmed with optical timing analysis alone. The significant reduction in optical flux, which is dominated by the disk and hotspot in IPs, supports the hypothesis that the disk mass significantly decreased during the low state. We note here that \citet{Ferrario-1999} modeled the power spectrum of intrinsic optical emission from IPs and acknowledged that a full model would also need to consider reprocessing of X-rays, though the model is still a useful reference when discussing accretion modes in IPs. 

\begin{figure*}
    \centering
    \includegraphics[width=0.95\textwidth]{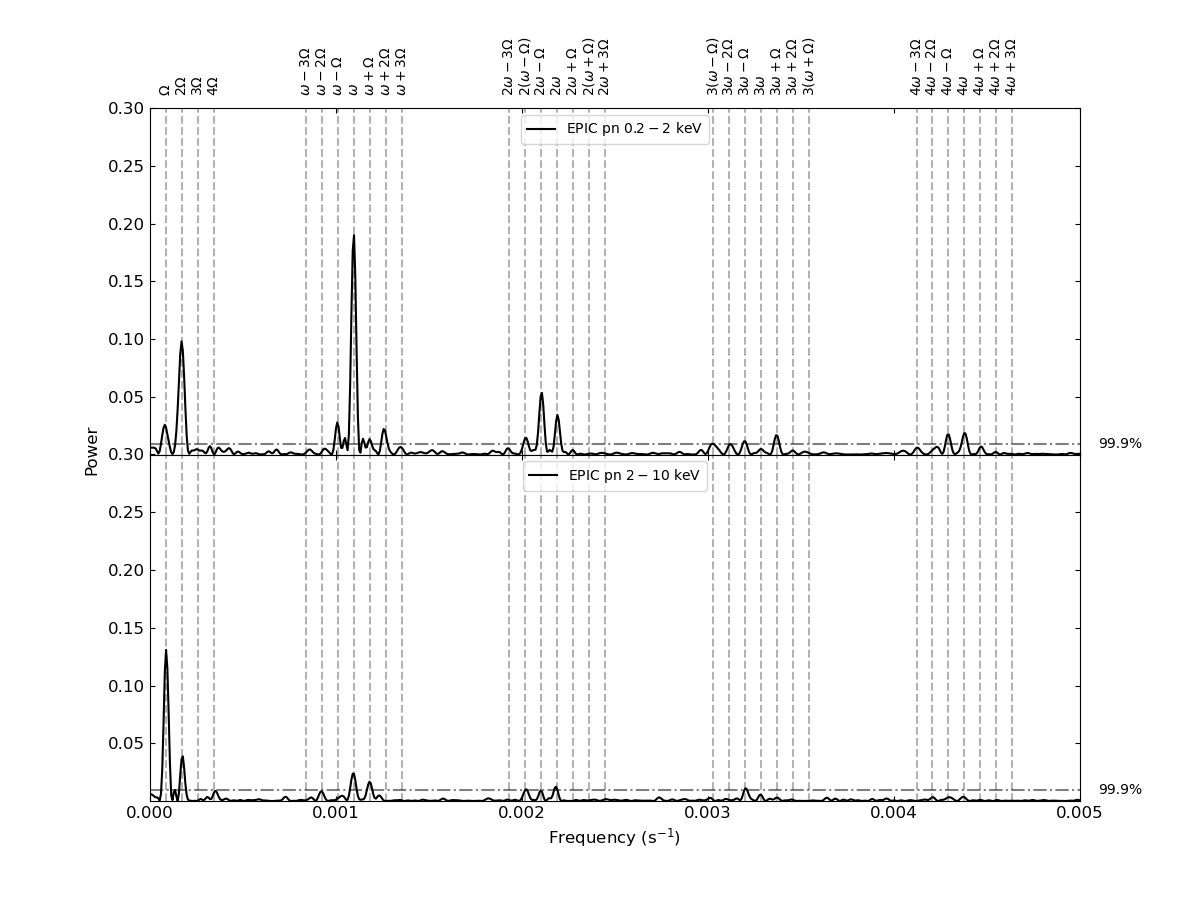}
    \caption{{\em Upper Panel}: Lomb-Scargle periodogram of the \xmm\ EPIC/pn 0.2--2 keV light curve of \src. {\em Lower Panel}: Lomb-Scargle periodogram of the EPIC/pn 2--10 keV light curve of \src. As in Fig. \ref{fig:XMM_LS}, vertical dashed lines represent known periodicities in the system.}
    \label{fig:energyres_LS}
\end{figure*}

However, the \citet{Ferrario-1999} model does not account for the increased variability at $2\Omega$ at X-ray energies that we see during the low state of \src. To resolve this, we consider the possibility that disk-overflow accretion might be a valid mechanism to explain this observed phenomenon. \citet{Hellier-1993} shows that the X-ray light curves of FO Aqr are compatible with a hybrid model of accretion in which accreting material flows both through a disk and a stream above the disk, known as disk-overflow. \citet{Hellier-1993} showed that the orbital modulation in the disk-overflow IP FO\,Aqr manifests in the form of two energy-dependent dips per orbit in the X-ray light curves. The energy dependence of the dips implies photoelectric absorption caused by increased scale height obscuring the X-ray emission at two locations: where the stream from the secondary hits the outer edge of the disk, and where the overflow stream hits the magnetosphere. To test if this mechanism is responsible for the $2\Omega$ peak in the X-ray power spectrum of \src\ during the low state, we generated Lomb-Scargle periodograms of the soft (0.2--2 keV) and hard (2--10 keV) \xmm\ light curves and plot them in Fig.\ \ref{fig:energyres_LS}. We find that the vast majority of the power at $2\Omega$ is found in the soft band and that the orbital variability in the hard band is almost all at $\Omega$. This is consistent with the \citet{Hellier-1993} disk-overflow accretion model and thus implies that \src\ is undergoing disk-overflow accretion during the low state, at least during the time of the \xmm\ observations.

We also consider the possibility of a double-armed spiral wave structure in the disk (generated by tides), which might generate two splash points where the arms meet the magnetosphere \citep{Murray-1999}, and thus variability in X-ray absorption at $2\Omega$. However, an argument against this scenario is that increased spiral-wave structure is thought to be more, not less, efficient at angular momentum transport, making the disk brighter, not fainter \citep{Murray-1999}. We thus prefer the disk-overflow accretion model.

\subsection{X-ray spectroscopy}

In Fig. \ref{fig:Swift_comp} we show the drastic difference in the \swift/XRT X-ray spectrum of \src\ between the normal high state and the low state. Contrary to what we see in the optical, during the low state, the observed (absorbed) X-ray flux of \src\ did not decrease relative to the high state. In fact, when examining the unabsorbed fluxes, we find a slight increase.

When modeling the X-ray spectra, the most striking difference we found was a significant reduction in absorption local to the system. When considering the model with two simple partial covering components (Model 1), we find that the covering fraction of the first {\tt pcfabs} component decreases from 93\% to 63\%, while the covering fraction of the second {\tt pcfabs} component decrease less drastically, from 67\% to 55\%. When utilizing a partial covering absorber combined with a complex absorber to represent absorption from the pre-shock flow ({\tt zxipab}; Model 2) we find that the covering fraction from the {\tt pcfabs component}, representing the absorption of X-ray photons by the disk, significantly decreases from 78\% to 30\%.

This significant decrease in covering fraction in both Model 1 and Model 2 is consistent with the hypothesis we presented in Section \ref{sec:discussion:timing}, i.e. that the disk had, at least partially, dissipated during the low state. Decreasing the mass of the disk in a moderately inclined source like \src\ decreases the likelihood of a photon being absorbed by the disk in the observer's line of sight.

The fact that the X-ray flux did not decrease between the typical high state low state implies that the mass-accretion rate on to the WD ($\dot{M}_{\rm WD}$) also did not decrease. This is contrary to what was seen during the 2016 low state of FO\,Aqr, which showed a dramatic decrease in X-ray flux relative to an archival high state observation, indicative of a reduction in $\dot{M}_{\rm WD}$ \citep{Kennedy-2017}. The earliest X-ray observation of \src\ during the 2025 low state took place on 2025-03-05 (MJD 60739), measuring $F_{\rm unabs, 0.5-10}=4.6^{+3.7}_{-1.1}\times10^{-11}$ erg cm$^{-2}$ s$^{-1}$ with \swift/XRT \citep{Shaw-2025}. This implies that $\dot{M}_{\rm WD}$ was comparable with the high state value from at least this date. However, there is no direct indication of $\dot{M}_{\rm WD}$ from the earliest part of the low state. We therefore cannot be certain that $\dot{M}_{\rm WD}$ ever decreased.

\subsection{Timeline of the low state of BG\,CMi}

Combining the multi-epoch, multi-wavelength timing data with the X-ray spectroscopy from both the high and low states of \src\ allows us to build up a near-complete picture of the physics of the low state. During the high state, \src\ can be considered to be a typical IP, in which the WD accretes matter via a partial accretion disk (the disk-fed accretion mode). However, in January 2025, the optical flux of \src\ decreased, indicative of the disk dissipating significantly. Though we are unable to determine whether or not $\dot{M}_{\rm WD}$ decreased at any point during the low state, the dissipation of the disk implies that the mass-transfer rate from the companion on to the disk  $\dot{M}_2<\dot{M}_{\rm WD}$, at least during the initial decline in optical flux (MJD $\sim60685-60699$; the orange portion of the light curve in Fig.\ \ref{fig:AAVSO_LC}). The dissipation of the accretion disk is not only supported by the drop in optical flux, but also by the optical periodograms in Fig.\ \ref{fig:AAVSO_LS}, which show a significant decrease in optical power at the spin frequency.

\citet{Covington-2022} note that at least one IP, DO\,Dra shows evidence of a complete cessation of accretion during a low state, with the white dwarf and its companion being the sole contributor to the optical flux. Therefore, we considered the possibility that the hot white dwarf contributed to the optical flux in the low state. \citet{Knigge-2011} show that the WD in \src\ ($P_{\rm orb}=3.23$ h) should have an effective temperature $T_{\rm eff}\sim25000$ K. The median absolute $V$-band magnitude for a WD in the range $20000-30000$ K is $M_V\sim10$ \citep{Tremblay-2011}. The measured distance to \src\ of $d=867^{+26}_{-24}$ pc \citep{Bailer-Jones-2021} implies that, even at its faintest, \src\ was $M_V\sim6$. We therefore do not consider the WD to make a noticeable contribution to the optical flux, even in the low state.

From MJD $\sim60751-60774$ (coincident with the light green portion of the optical light curve in Fig.\ \ref{fig:AAVSO_LC}), the optical flux of \src\ increased, back towards its typical level. At this point in  its evolution, the recovering optical flux implies that the disk was re-forming, and thus $\dot{M}_2>\dot{M}_{\rm WD}$. It was during this epoch that the \xmm\ observations of \src\ took place. Our (X-ray) spectral and (multi-wavelength) timing analysis indicates that accretion on to the WD was occurring via disk-overflow. The disk, though in the process of recovering, was still lower-mass than during the typical high state, and thus was less effective at absorbing X-rays. \citet{Lubow-1989} noted that the scale height of the ballistic stream from the companion on to the disk is typically bigger than that of the disk by a factor of 2--3. 
Overflowing material can then pass over the disk before being reincorporated into the rest of the disk \citep[see also][for example]{Armitage-1996,Armitage-1998}. We suggest here that the disk in \src\ was small enough that overflowing material was able to pass completely over the disk and reach the magnetospheric radius, $R_{\rm mag}$, where it then flowed along the field lines on to the WD. This is similar to what was seen during the low state of V515\,And, which went through a disk-overflow phase, before switching to stream-only accretion after the disk had fully emptied \citep{Covington-2022}.

After the optical flux of \src\ had recovered to its normal high state value (the dark green portion of the optical light curve in Fig.\ \ref{fig:AAVSO_LC}), we see the optical periodogram return to spin-dominated, much like it was prior to the low state. We interpret this as returning to a disk-fed accretion mode.


\section{Conclusions}
\label{sec:conclusions}

We have presented a multi-wavelength observing campaign of the 2025 low state of the IP \src. Optical monitoring of the source show that its flux dropped by $\sim0.5$ mag for $\sim 50$ d. We interpret this as being due to the mass transfer rate from the companion on to the disk, $\dot{M}_2$, decreasing to a point where it was less than the rate at which mass was transferred from the disk on to the WD ($\dot{M}_2<\dot{M}_{\rm WD}$). This caused the disk to empty out over the course of the low state, though we cannot be certain that the disk ever completely dissipated. X-ray observations close to the end of the low state reveal a spectrum with significantly lower absorption than archival \suz\ observations of the source, which we interpret as the disk being lower mass than typical, and thus being a less-effective absorber of X-rays in our line of sight. We also find strong evidence for disk-overflow accretion during this portion of the low state, something which has been seen in numerous other IPs during their low states \citep[see e.g.][]{Kennedy-2017,Littlefield-2020,Covington-2022}. However, we do not find strong evidence for a complete switch to a stream-fed accretion mode.

We still do not have strong, testable theories for why $\dot{M}_2$ occasionally decreases in IPs, leading to low states. Recent studies of the other class of magnetic CVs, polars, have revealed evidence of a magnetic valve regulating mass transfer and thus controlling the high-low state cycle \citep{Mason-2022,Mason-2024}, but the magnetic fields of the WDs in IPs are significantly weaker and thus not expected to influence mass transfer at $L_1$. \citet{Livio-1994} proposed that starspots on the companion passing in front of the $L_1$ point are the cause of the observed drop in the optical flux of VY Scl stars, and thus, such a mechanism could theoretically be extended to IPs. \citet{Bianchini-1990} suggested that late-type donors in CVs can exhibit fractional variations in stellar radius. Any variation in donor radius could potentially affect $\dot{M}_2$ due to the Roche Lobe overflow nature of IPs. However, it remains to be seen whether any, or none, of the above scenarios are responsible for the low states in IPs. 

\begin{acknowledgments}

The authors would like to thank the anonymous referee for providing useful feedback which helped improve the manuscript. AWS would like to thank Professor Ed Cackett for contributing observations to the BG\,CMi campaign with the Dan Zowada Memorial Observatory. 
We acknowledge with thanks the variable star observations from the AAVSO International Database contributed by observers worldwide and used in this research. We thank Norbert Schartel and the {\em XMM-Newton} OTAC chair for granting ToO observations of BG\,CMi at short notice. This work made use of Astropy,\footnote{http://www.astropy.org} a community-developed core Python package and an ecosystem of tools and resources for astronomy \citep{Astropy-2013,Astropy-2018,Astropy-2022}. This research made use of {\tt ccdproc}, an Astropy package for image reduction \citep{Craig-2017}. This work was supported by Slovak Research and Development Agency under contract No. APVV-20-0148. This article is based on observations made with the IAC80 telescope operated on the island of Tenerife by the Instituto de Astrof\'{i}sica de Canarias in the Spanish Observatorio del Teide. COH is supported by NSERC Discovery Grant RGPIN-2023-04264, and Alberta Innovates Advance Program 242506334. CGN was supported by the Butler Summer Institute.

\end{acknowledgments} 

\facilities{XMM, Suzaku, Swift, OT:0.8m, Elizabeth, Csere, ING:Kapteyn, Great Basin, AAVSO}

\bibliography{references}{}
\bibliographystyle{aasjournalv7}



\end{document}